\documentstyle[12pt]{article}
\input epsf
\begin{document}

\begin{titlepage}

\begin{tabular}{l}
May, 1996                   
\end{tabular}
    \hfill
\begin{tabular}{l}
MSU-HEP-50812 \\
FSU-HEP-951031 \\
CTEQ-512 \\
\end{tabular}

\vspace{2cm}

\begin{center}

\renewcommand{\thefootnote}{\fnsymbol{footnote}}

{\LARGE Large Transverse Momentum Jet 
Production and the Gluon Distribution Inside the Proton\footnote[1]{
This work was supported in part by the DOE and NSF.}}

\renewcommand{\thefootnote}{\arabic{footnote}}

\vspace{1.25cm}


{\large J.~Huston$^d$, E.~Kovacs$^b$, S.~Kuhlmann$^a$, H.~L.~Lai$^d$,
J.~F.~Owens$^c$, D.~Soper$^e$, and W.~K.~Tung$^d$}

\vspace{1.25cm}

$^a$Argonne National Laboratory, \\
$^b$Fermi National Accelerator Laboratory, \\
$^c$Florida State University, \\
$^d$Michigan State University, \\
$^e$University of Oregon 

\end{center}
\vfill

\begin{abstract}
The CDF experiment has reported an excess 
of high-$p_t$ jets compared to previous next-to-leading order 
QCD expectations. 
Before attributing this to new physics effects, we investigate
whether these high-$p_t$ jets can be explained by a modified gluon
distribution inside the proton.  
We find enough flexibility in a global QCD analysis including the CDF inclusive
jet data to provide a 25-35\% increase in the 
jet cross sections at the highest $p_t$
of the experiment. Two possible sets of parton distributions are presented, 
and the effects of these on other existing data sets are presented.
Further theoretical and experimental work needed to clarify unresolved 
issues is outlined.

\end{abstract}

\vfill
\newpage
\end{titlepage}

\begin{figure}[tbp]
\begin{center}
\begin{minipage}[h]{6.5in}
\epsfxsize=6.3in
\epsfbox[36 144 520 650]{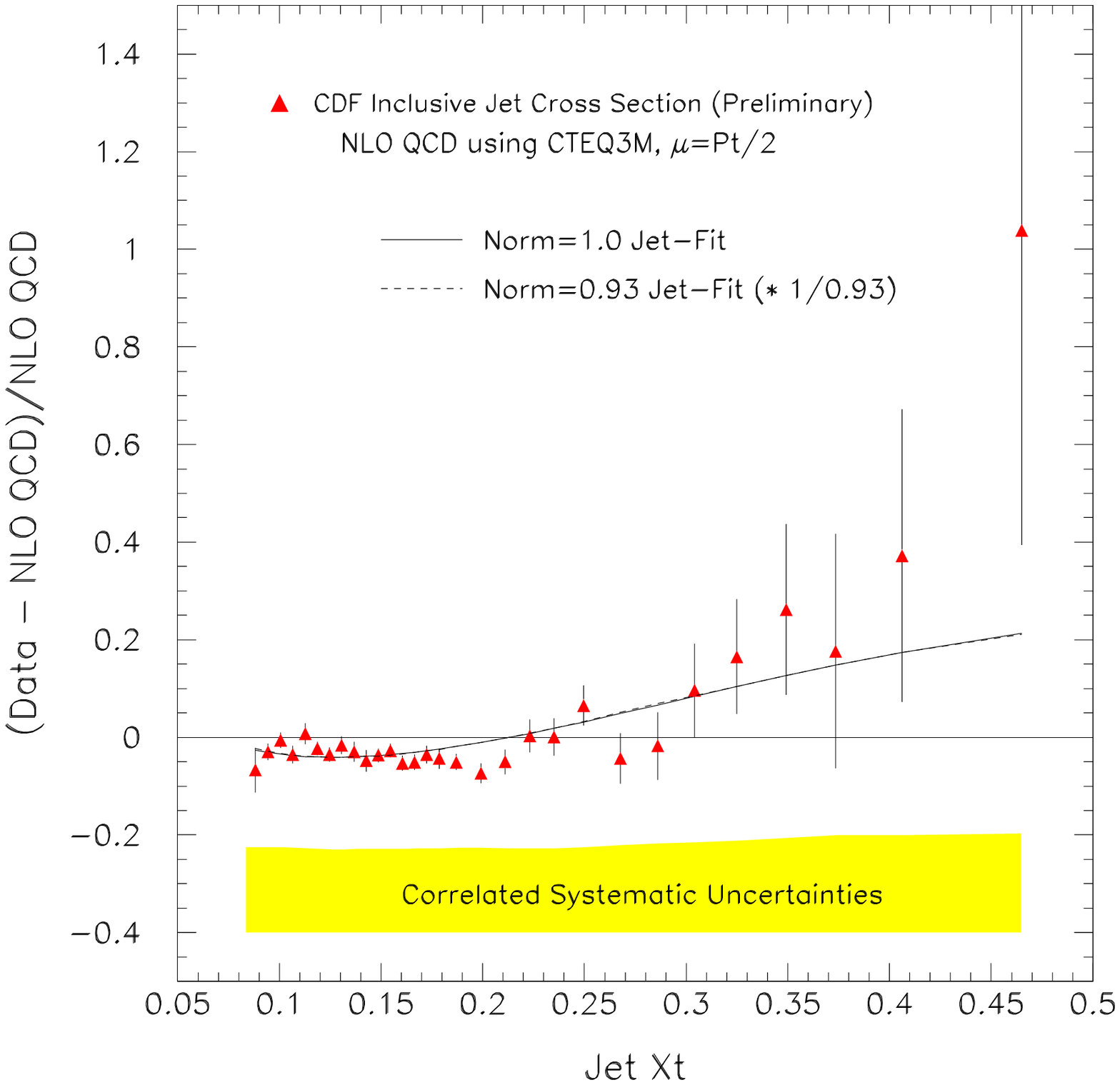}
\caption{The preliminary CDF jet data is compared to a 
NLO QCD calculation using the conventional CTEQ3M parton distributions 
(points), and the new parton distributions fit to the jet data (solid 
and dashed lines that lie on top of each other).}
\label{jf3341}
\end{minipage}
\end{center}
\end{figure}

Jet production in hadron collisions at the Fermilab Tevatron is an important
process which presently provides the highest energy for studying hard
scattering dynamics. 
Quark substructure or other new short distance physics would, if present, 
change the cross section for high-$p_t$ jet production~\cite{ehlq}.
Such a deviation compared to next-to-leading order (NLO)
perturbative Quantum Chromodynamics (QCD)~calculations, based
on commonly used parton distributions, has been reported by the CDF 
experiment~\cite{anwar} in the range $200<p_t<420$ GeV from 20 $pb^{-1}$ 
of data.  These data are shown in Fig.~\ref{jf3341}, indicating a 
clear 40\% excess at $p_t=350$ GeV compared to the NLO 
calculation~\cite{eks}.
The points are (Data-NLO~QCD)/NLO~QCD plotted versus the 
scaling variable $x_t=2p_t/\sqrt{s}$.  The
theory is calculated with CTEQ3M parton distributions~\cite{cteq3} and 
$\mu=p_t/2$.
The region of excess corresponds to $0.22<x_t<0.45$.  As more
data are analysed, this excess could become more significant, as
suggested by the first results on the dijet mass distribution 
from 70~$pb^{-1}$ of data~\cite{anwar}.
In order to determine whether this enhancement constitutes a signal for 
{\it new physics}, it is crucial to investigate possible 
explanations within the Standard Model.  

One well-known uncertainty concerns the dependence of perturbative
calculations on the choices of the renormalization and factorization scales.
However, for inclusive jet cross sections, this dependence is 
quite small (10\%) and is largely independent of $p_t$~\cite{eks}.
Similarly, changes in the 
strong coupling $\alpha_s$ resulting from variations of 
$\Lambda_{QCD}$ (such as a comparison between calculations using CTEQ2M
and CTEQ2ML~\cite{cteq3}, or the recent MRS\cite{mrslam} parton
distributions) mainly affect the normalization.
Another source of uncertainty is the effect of summing large perturbative
logarithms that may be important at large-$x_t$ and 
have been shown to be significant for high-mass lepton-pair
production\cite{sterman}. A corresponding study for jet cross sections has not
yet been carried out. 
In addition, the long-standing disagreement
between NLO QCD and the jet $x_t$ scaling result from CDF~\cite{xtscale} 
points to a potential inadequacy in the NLO calculations, or a 
possible mismatch between the theoretical and experimental 
jet definitions.
However, it is not clear whether this effect, even if it is real, will
extend to the $x_t$-region under consideration.

Finally, there are the parton distributions which play a crucial role in
determining the perturbative QCD ``predictions'' of the jet cross section.
For the $x_t$ range in question, more than 50\% of the jet cross section is
due to quark-quark scattering, and the quark distributions are well
determined by the precise data from deep inelastic lepton-nucleon scattering
(DIS). On the other hand, although the gluon distribution is small in this
region, its contribution to the cross section (mainly through the
gluon-quark scattering processes) is still substantial---of the order of
25-50\%. The DIS data do not constrain the gluon much at large $x$, that
role being usually played by direct photon production data in most modern
global analyses.\cite{cteq3,mrsa} In the light of current theoretical and
experimental uncertainties on direct photon production, it remains an open
question whether the usual gluon distributions can be modified in the
relevant $x$ region to accommodate the observed high-$p_t$ jets. 
The purpose
of this paper is to report on a quantitative study addressed to this
particular question.

\paragraph{Global QCD Analysis Incorporating CDF Jet Data}
We have carried out a global QCD analysis~\cite{cteq3} 
incorporating for the first time the CDF inclusive jet data.
Particular attention is given to the compatibility of the inclusive
jet data with the collection of data sets used in previous global analyses
within the NLO QCD framework \cite{cteq3,mrsa}. 

As seen in Fig.~\ref{jf3341}, jet production data with $p_t>200$ GeV 
($x_t>0.22$), where the excess occurs, have comparable experimental
systematic and statistical errors, whereas below this value the systematic
errors dominate. Thus, these two regions are given separate attention. For
the lower $p_t$\ range, although the 
experimental measurements extend down to $p_t=15$~GeV, we chose 
to include only data with $p_t>75$~GeV in this
study due to a number of potential theoretical and experimental problems
relating to low-$p_t$ jets. These include: 1) possible problems in the match
between theory and experimental jet definitions, such as fragmentation products
outside the jet cone, 2) definitions of the ``underlying event'' coming from
the proton-antiproton remnants, 3) scale uncertainty of NLO QCD calculations
which becomes non-negligible at low $p_t,$ and 4) $k_t$ broadening
(discussed later for direct photons). All of these affect low-$p_t$ jets
much more than high-$p_t$ jets. 

Our systematic study reveals that there is enough flexibility in the NLO
QCD global analysis framework to enhance the theoretical cross section for
the highest $p_t$ inclusive jets by 25-35\% above the previous calculations. We
will describe two sample parton distribution sets~\cite{whereglu} 
which illustrate two
slightly different ways that the overall fit can be accomplished.  The first,
designated as the norm=1.0 jet-fit, fixes the CDF jet data normalization at
the nominal value so that the high-$p_t$ excess points are
accommodated without an overall downward shift of experimental points.
However, without fixing the normalization of the jet data, the global analysis
prefers a relative downward shift of the CDF data with respect to theory. 
The second example, the norm=0.93 jet-fit, is chosen to represent this
possibility.  Both solutions give good fits to the other data sets
included in the global analysis.

Fig.~\ref{jf3341} includes two curves corresponding to NLO QCD calculations
using parton distributions from the two new fits along with the CDF jet
data: the solid line for the norm=1.0 jet-fit, and the dashed line for the
norm=0.93 jet-fit(divided by 0.93). The two new fits lie virtually on top
of each other.  
The total $\chi^2$ for the 1147 DIS, Drell-Yan, direct photon, and 
CDF jet data points in the norm=1.0(0.93) jet-fit is 1160(1130), 
clearly quite good.
Both of the new fits remove much of the excess of the
large $p_t$ jet data, with a $\chi^2/\#pts = 1.36$, which is quite 
acceptable considering this ignores the
systematic uncertainties in the jet data.  
The quadratic sum of eight different CDF systematic uncertainties is 
shown as a shaded band below the data points.  While the size
of the band appears independent of jet $p_t$, the eight individual 
uncertainties are not; they must be folded in for
a proper analysis of errors after detailed information becomes available
from CDF.  But for our purposes of determining if the jet data 
can be accommodated within QCD uncertainties, the proper procedure
is to only fit the jets with statistical uncertainties and give 
this data set more weight in the global fit, then look closely 
at the other data sets in the fit to see if discrepancies arise.
This does not imply that one obtains the best estimate of the true parton 
distributions in nature; it does prove that viable parton sets exist.
The detailed comparison of the jet-fit partons with other data sets will
be described later.

\begin{figure}[tbp]
\begin{center}
\begin{minipage}[h]{6.5in}
\epsfxsize=6.3in
\epsfbox[36 144 520 650]{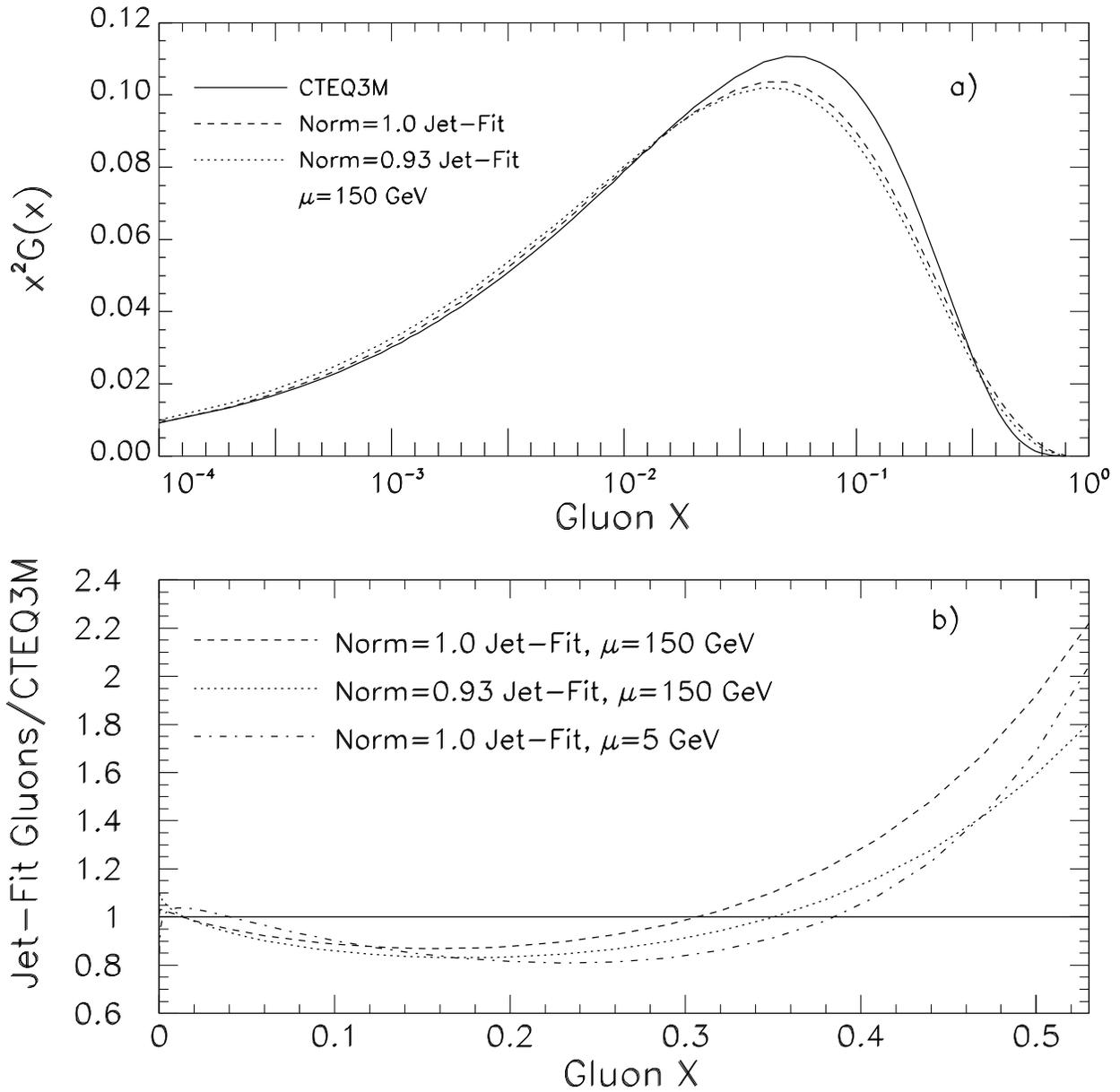}
	\caption{(a) The gluon distributions at $\mu=150$ GeV
from the norm=1.0 and the norm=0.93 jet-fits are compared to that of CTEQ3M:
(b) the ratio of the two jet-fit gluons to CTEQ3M (see text).}
	\label{higl150}
\end{minipage}
\end{center}
\end{figure}

The gluon distributions from the two new fits are compared with that of 
CTEQ3M in Fig.~\ref{higl150} at $\mu=150$ GeV, which corresponds to the 
middle of the high-$p_t$ data range with $\mu=p_t/2$.
In Fig.~\ref{higl150}a, $x^2G(x)$ is plotted against $\log x.$
(Since $x^2f(x)\cdot d\log x=xf(x)\cdot dx$ is the 
momentum fraction within $dx$,
each curve in this plot directly depicts the distribution of momentum
fraction carried by the gluon.)  In Fig.~\ref{higl150}b, the ratio of the
jet-fit gluons to that of CTEQ3M is shown over the $x>0.1$ range. For the
norm=1.0 fit, we see a significantly increased $G(x)$ in the large-$x$
region, with a compensating decrease in the medium-$x$ region and little change
in the lower-$x$ range. 
For the norm=0.93 fit, $G(x)$ is uniformly shifted down from the 
norm=1.0 fit in the range $0.05<x$, with a compensating increase
in the small-$x$ region. This shows that the jet data used in the fit 
constrain the {\it shape} of $G(x)$ in the region $0.08<x<0.45$.
Also shown in Fig.~\ref{higl150}b, is the
ratio for $\mu=5$ GeV, which is relevant for discussing comparisons with direct
photon data later.
We also note that the $\alpha_s(M_Z)$ values for the new fits are 
slightly higher, $0.116$ compared to $0.112$ for CTEQ3M.

\paragraph{Comparisons to Deep-Inelastic Scattering and Direct Photon Data}
Deep-inelastic scattering data are indirectly sensitive to the gluon
distribution through NLO corrections and scaling violations. But at 
large-$x$ the effects on $F_2$ due to a modified gluon 
distribution can be easily compensated by small
changes in the quark distributions and in $\Lambda_{QCD}$. 
A detailed look at the shifts in $F_2$ for
the various DIS experiments shows no changes of more than 2\% between CTEQ3
and the jet-fit results for all values of $x$ and $Q^2$. 

Fixed target direct photon data have usually been regarded as the main
source of constraint on the gluon distribution at large-$x$. However, the
constraint is weakened if the theoretical uncertainties unrelated to
parton distributions are significant.
We will now review the relevant theoretical issues, then
evaluate the effect of the jet-fit gluons in light of these uncertainties.
For definiteness, we shall use the most widely used WA70 data as the point
of discussion, although the results are independent of the specific
experiment.

The two most significant theoretical uncertainties for fixed target direct
photons are the factorization scale dependence, and the possible $k_t$
broadening effect. The latter is suggested by a recent global study of all
direct photon data~\cite{phokt}; it involves the likelihood that NLO QCD
does not contain enough of the $k_t$ of the initial state gluon radiation,
thus leading to an underestimate of low-$p_t$ photon 
cross sections. Since the publication
of Ref.~\cite{phokt}, there have been two developments which further support
the basic idea of $k_t$ broadening in direct photon production: 1) a new
calculation of collider direct photon production incorporating NLO QCD hard
scattering plus initial state parton showers~\cite{howie} 
shows good agreement with the
shape of the CDF direct photon data~\cite{cdfph}. 2) The preliminary,
high-statistics, E706 direct photon data~\cite{e706} (the most precise
measurement yet at fixed target energies) also shows a significant excess of
photons compared with NLO QCD calculations. This excess is largest when the $%
p_t$ slope of the data is greatest, which again is consistent with the
expectations from a $k_t$ broadening effect.

\begin{figure}[tbp]
\begin{center}
\begin{minipage}[h]{6.5in}
\epsfxsize=6.3in
\epsfbox[36 144 520 650]{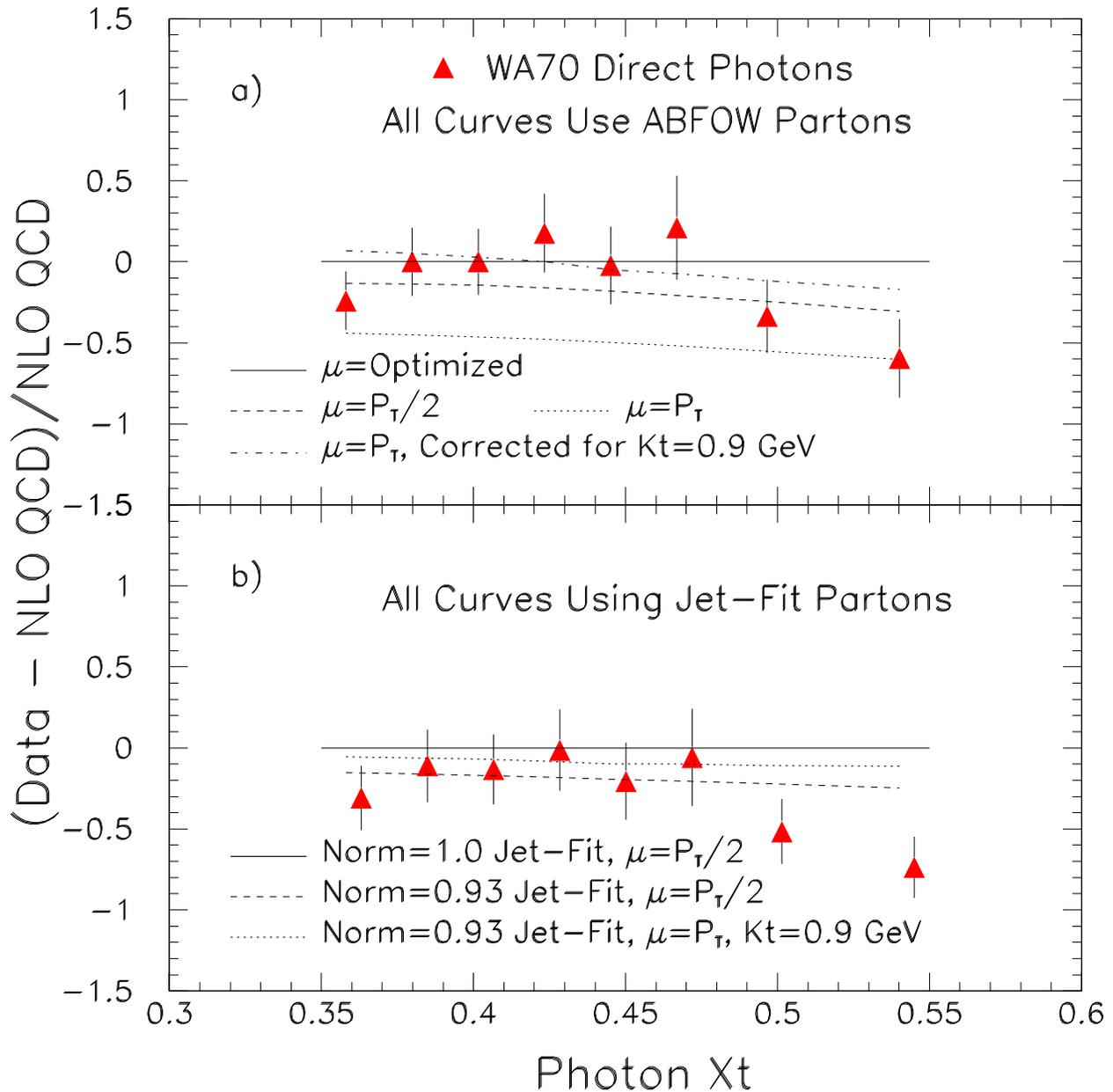}
	\caption{The WA70 direct photon data is compared to 
NLO QCD calculations using conventional, ABFOW parton
distributions in a). Different choices of scale are shown as well
as the effect of adding additional $k_t$ broadening to the
theory.  In b) the WA70 direct photon data is compared to 
NLO QCD calculations using the two sets of jet-fit gluons.
(see text)}
	\label{wa70_2}
\end{minipage}
\end{center}
\end{figure}

In Fig.~\ref{wa70_2}a, the WA70 direct photon
data is compared to NLO QCD for a variety of scales, using conventional
ABFOW parton distributions~\cite{abfow}. The change in theoretical value in
going from optimized $\mu $\ (used by ABFOW and MRS) 
to $\mu =p_t$ is about 50\%. This large
variation due to scale changes provides a measure of the theoretical
uncertainties due to higher order corrections. 
Next, to show the effect due
to a possible $k_t$ broadening, we also include in Fig.~\ref{wa70_2}a a curve
corresponding to a scale choice of $\mu =p_t$ plus an average $k_t$
broadening of $0.9$ GeV using the algorithm of reference~\cite{jeffrev}. 
The number $0.9$ GeV comes from the WA70 analysis
of their diphoton measurement~\cite{wa70di}. We see that the broadening
correction is also about 50\%, and brings the $\mu=p_t$ curve into
agreement with the data.

Even with the large theoretical uncertainties described above, one might
still expect the fixed target data to rule out one or both of the jet-fit
gluons because, naively, the differences between the jet-fit and
conventional gluons at a typical $\mu=150$ GeV might be significantly
amplified at the low $\mu$ value (2 GeV to 6 GeV) of the fixed-target
experiments. But this is not the case: the crossing 
point between the
jet-fit and CTEQ3M gluons occurs around $x\approx 0.4$ at $\mu=5$ GeV, 
as shown in Fig.~\ref{higl150}b. The comparison of the
jet-fit results with the WA70 direct photon data is shown in 
Fig.~\ref{wa70_2}b.  In the solid and dashed curves we have used a
scale of $\mu =p_t/2$ with no $k_t$ corrections.  In the dotted 
curve we used $\mu=p_t$ and a $k_t$ broadening of 0.9 GeV.
All three curves are consistent with the WA70 data which has
a 10\% normalization uncertainty.
These results clearly demonstrate that given the uncertainties 
with scale choice and $k_t$ broadening, the new gluon distributions
are fully consistent with the WA70 data.  As mentioned above, similar 
results hold for other fixed target direct photon data sets.

\paragraph{UA2 Inclusive Jet Data}

Of considerable interest to our study of high-$x_t$ jets is the earlier UA2
inclusive jet cross section~\cite{ua2jet} measurement. The data have high
statistics, are in the same $x$ range as the CDF measurement, and cover a
similar rapidity range.  
Although the two experiments are at different scales set by the
respective $p_t$ ranges, the QCD evolution between the two is
not significant, hence they essentially probe the 
same parton distributions.
There are some important differences however. For
the same $x_t$, the UA2 jets are at lower $p_t$, and may be
subject to the additional low-$p_t$ uncertainties that were
discussed above.
In addition, the UA2 data are based on a jet finding algorithm
that less closely follows the infrared-safe ``Snowmass'' 
algorithm~\cite{snow}, in fact the UA2 publication itself
expresses caution concerning comparisons with NLO QCD.
In our NLO theory calculations for CDF we use the Snowmass algorithm
with $R=0.7$ at the parton level, while we model the UA2 algorithm
with the modified Snowmass algorithm~\cite{eks2} with $R=R_{sep}=1.37$.
Both the low-$p_t$ and jet algorithm effects warrant further 
study.  To account for them at present, we would assign a 
larger theoretical uncertainty, 20\% compared to the nominal 10\%, 
for the UA2 calculations.

\begin{figure}[tbp]
\begin{center}
\begin{minipage}[h]{6.5in}
\epsfxsize=6.3in
\epsfbox[36 144 520 650]{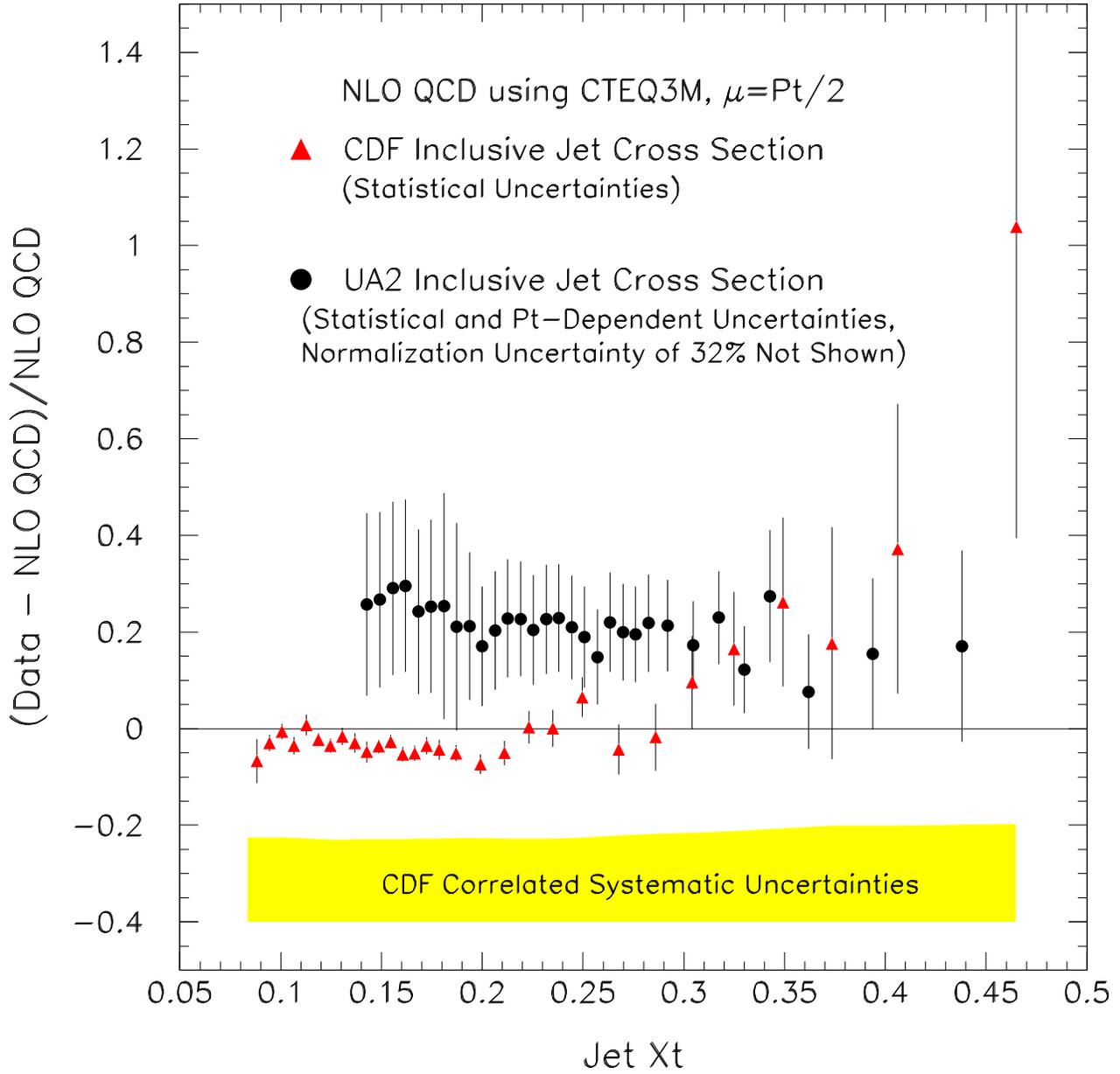}
	\caption{The CDF and UA2 jet production measurements are 
compared to NLO QCD calculations (see text).}
	\label{ua2cdfct3m}
\end{minipage}
\end{center}
\end{figure}

Fig.~\ref{ua2cdfct3m} shows the CDF and UA2 jet data compared to these NLO QCD
calculations using CTEQ3M parton distributions, and $\mu=p_t/2$. 
The CDF data points have statistical uncertainties only, 
while the UA2 points include statistical and $p_t$-dependent 
systematic uncertainties (this is the way the two different groups
present their data). There is an additional 32\% normalization uncertainty
in the UA2 measurement, while the CDF correlated systematic uncertainty 
band is shown at the bottom of the plot.  The UA2 data are 
systematically larger than the theory (but within the normalization 
uncertainty), but in general there is no distinct shape difference 
as is seen in the CDF data.  If one ignores experimental
uncertainties, the two experiments disagree with each other.  But
clearly both experiments have complicated correlated systematic 
uncertainties that need to be understood before 
conclusions can be drawn.  

\paragraph{Conclusions}

If the excess of high-$p_t$ jets at CDF persists, it will be one of the
most important challenges for QCD. Understanding what role experimental
uncertainties and/or conventional theoretical sources play is
crucial to understanding whether there is new physics present. This paper
has considered in detail only one possible conventional theoretical source,
parton distributions, especially the gluon distribution. 
Two examples were given which show that there is
considerable room to modify the gluon distribution so that current inclusive
jet data can be incorporated in an overall global NLO QCD fit. On the other
hand, taken at face value, the UA2 jet measurement disagrees with the CDF
data and prefers the shape of conventional gluon 
distributions such as CTEQ3M. This disagreement may or may 
not be accounted for by experimental systematic uncertainties. 

It will very likely take many years and much work to resolve these issues.
Beyond the obvious need for an independent, robust, method of measuring
high-$x$ gluons, we list five potential studies/calculations that
are needed to understand the excess: 
1) measuring the dijet angular distribution as an independent QCD test
that differentiates between parton distributions and new interactions;
2) incorporating the correlated systematic
uncertainties from CDF and UA2 jet measurements into the global analysis to
determine if they are compatible; 
3) confirming and understanding the jet $x_t$ scaling result from CDF with a
new lower energy run at Fermilab; 
4) performing the large-$x$ resummation
for jet production as has been done 
for the Drell-Yan process, to see if the excess
can be so explained; and 5) performing a $p_t$ resummation of soft
gluons in the direct photon calculation to reduce the
uncertainties due to the scale dependence and $k_t$ broadening, hence
sharpening the constraints on the gluon distribution due to these two
complementary processes. Before one can claim there is {\em new physics} in
the CDF jet excess, it is likely that all five of these future studies will
be necessary.

Note added in proof:  A recent paper from the GMRS group~\cite{gmrs}
has stated it is impossible to obtain parton distributions 
in agreement with the CDF jet data.  Their attempt to modify 
the quark distributions to fit the jet data results in a 
$\chi^2$ of 20703 for 128 BCDMS data points.  In contrast, 
the two jet-fits presented in this paper give rise to a 
$\chi^2$ of 173 and 175 respectively for 168 BCDMS data points.
It appears the reason that GMRS were not able to find 
satisfactory solutions like ours is that they did not allow
sufficient flexibility in the gluon distribution shape.

\newpage


\clearpage

\clearpage

\clearpage

\clearpage

\end{document}